\documentclass [reqno]{article}
\title{\vspace{-25pt} Response to \emph{Refutation of Aslam's Proof that }\textbf{NP = P}\\
}

\author{
Javaid Aslam\footnote{\copyright Copyright Javaid Aslam 2009-2017, Santa Clara, CA 95054}
\\
jaslamx@yahoo.com
}

\usepackage{wrapfig}
\usepackage{float}
\usepackage{fancyhdr}
\usepackage{subfig}
\usepackage{calc}
\usepackage{amsthm}
\usepackage{amsfonts, stmaryrd}
\usepackage{amssymb, amsmath, latexsym}
\usepackage[utf8]{inputenc}
\usepackage[T1]{fontenc}
\usepackage{lmodern} 
\usepackage{setspace}
\usepackage[usenames]{color}
\usepackage{url}

\usepackage{lineno,hyperref}
\usepackage{algorithmic}
\usepackage[section]{algorithm}
\usepackage{eqparbox}

\usepackage[titletoc, title]{appendix}
\usepackage{graphicx}
\newcommand{\defn}{\overset{ \text{def}}{=}}
\newcommand{\less} {\textless{}}
\newcommand{\more} {\textgreater{}}
\newcommand{\vmpset}{%
V\hspace{-2pt}M\hspace{-1.50pt}P\hspace{-0.5pt}Set}%
\renewenvironment{proof}{{\bfseries Proof.}}{\par \qed \par}
\newcommand{\cvmp}{CV\hspace{-2pt}M\hspace{-1.0pt}P}
\newcommand{\vmp}{V\hspace{-2pt}M\hspace{-1.0pt}P}
\newcommand{\cms}{C\hspace{-1pt}M\hspace{-1.0pt}S\,}
\newcommand{\gms}{G\hspace{-1pt}M\hspace{-1.0pt}S\,}
\newtheorem{theorem}{Theorem}[section]

\newtheorem{lemma}[theorem]{Lemma}
\newtheorem{claim}[theorem]{Claim}

\theoremstyle{definition}
\newtheorem{definition}[theorem]{Definition}
\theoremstyle{remark}
\numberwithin{equation}{section}

\renewcommand{\abovecaptionskip}{7pt}
\renewcommand{\belowcaptionskip}{5pt}
\begin{document}
\vspace{-40pt}
\maketitle
\raggedright
\vspace{-15pt}
\parindent 0.0in
\begin{abstract}
 This paper provides a further refinement to the previous response by introducing new structures and algorithms for counting VMPs of common
 \emph{Edge Requirement} (ER) and hence for counting the perfect matchings.
 \end{abstract}
\vspace{-10pt}
\section{The ER Satisfiability and Enumeration}
\vspace{-10pt}
 The distinguishing result of the main paper \cite{aslam-2017} is a P-time enumerable \emph{partition} of all the potential perfect matchings in a bipartite graph. This partition is a set of equivalence classes induced by the missing edges in the potential perfect matchings. We capture the behavior of these missing edges in a polynomially bounded representation  by a graph theoretic structure, called MinSet Sequence, where MinSet is a P-time enumerable structure derived from a graph theoretic counterpart of a generating set of the symmetric group.\par
 The above  MinSet Sequences can be viewed  as a transformed  problem of Perfect Matching, and which can be summarized by some of its characteristic attributes as follows:
\vspace{-6pt}
\begin{itemize}
\item 	The generators of MinSet Sequences are created in the following two main steps:
\vspace{-4pt}
\begin{enumerate}
\item 	Partition and transform the Cosets of the symmetric group $S_n$  into disjoint subsets, called  CVMPSet, containing perfect matchings from $K_{n,n}$.\vskip 3pt
\par
     To achieve this, we first map a specific generating set  of the symmetric group $S_n$  to a set of graph theoretic generators, for generating all the perfect matchings (PM) in $K_{n,n}$. This means mapping each set of (right) coset representatives $U_i$ to a graph theoretic counterpart, called \emph{partition representatives} $g(i)$. Two kinds of binary relations over $\{g(i)\}$ model the multiplicative behavior of these generators, leading to  a generating graph, $\Gamma(n)$ (a directed $n$-partite graph), for generating all the PMs in $K_{n,n}$, where each PM is represented by a directed path in $\Gamma(n)$, called CVMP, which is a sequence of $n$ unique generators from $g(1)\times g(2)\times \cdots\, \times g(n-1) \times g(n)$. A VMP is any sub-path of a CVMP.
Now each graph theoretic ``coset representative'' (a generator from $g(1)$)  induces an equivalence class over each Coset, called CVMPSet.
\begin{figure}[h]
\vskip -9pt
\vspace{ -05pt}
\center{
\includegraphics[scale=0.50]{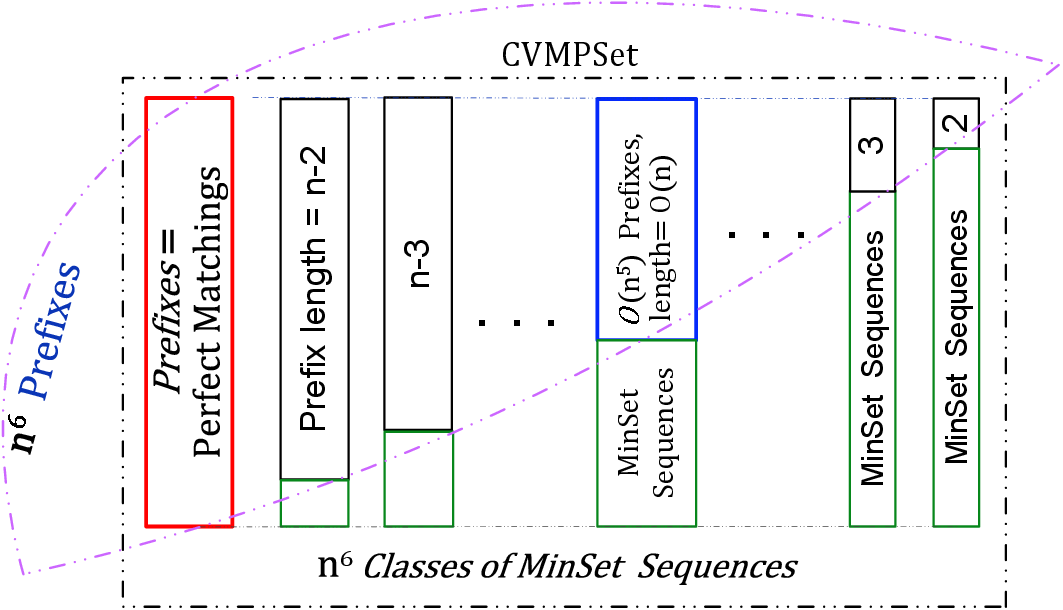}
\vskip -5pt
\caption{\textbf{A Partition Containing MinSet Sequences}}\label{FIG:cosetPartitions}
}
\end{figure}
\item Partition each CVMPSet into MinSet Sequences, where each sequence consists of  a small subset of the  polynomially many MinSets defined as follows.
    \par


\par
For any bipartite graph BG, CVMPs represent only \underline{potential} perfect matchings, and therefore, they are qualified by an attribute called Edge Requirement (ER) defined as follows. ER of a CVMP contains all the missing edges in the potential PM represented by that CVMP. A MinSet contains all the VMPs of common ER, with the missing edges only at three distinguished nodes on the VMPs. A judicious choice of the common nodes of these VMPs in a CVMPSet allows a MinSet and any sequence of the MinSets to be P-time enumerable.
Each CVMPSet can thus be decomposed into disjoint subsets, each being a (set of a) unique sequence of MinSets, representing a disjoint subset of the $n!$ potential perfect matchings in BG.\\

While there can be exponentially many MinSet sequences, there are only polynomially many classes induced by the polynomially many ($O(n^6)$) prefixes of the MinSet sequences. Only a sequence containing exactly one MinSet can contain perfect matchings in BG.

\end{enumerate}
\item Each sequence of length 1 (containing exactly one MinSet), containing CVMPs of length $n-1$ and with null ER, captures a disjoint subset of  the
 perfect matchings in BG.\\ When $BG  =  K_{n,n}$,  $MinSet  =  CVMPSet$.\\
Other MinSet sequences can also be enumerated, but they are not of interest.
\end{itemize}

\vspace{-10pt}
\vspace{-0pt}
\vspace{-0pt}
\section{Key Concepts}
\vspace{-11pt}
Let $G^{(i)}$ be a subgroup of a permutation group $G < S_n$, obtained from $G$ by fixing all the points in $\lbrace 1, 2, \cdots, i \rbrace$, where $1 \le i \le n$. That is, $\forall\,\pi \in G^{(i)}$, and $\forall j \in \lbrace 1, 2, \cdots, i \rbrace$, $~j^\pi = j$. Then  $G^{(i)} < G^{(i-1)}$,  $G^{(0)} = G$.
Then the following sequence of subgroups is referred to as a \emph{stabilizer chain} of $G$.
\vspace{-7pt}
\begin{equation}\label{e:tower}
I = G^{(n)} < G^{(n-1)} < ~ \cdots ~ <G^{(1)} < G^{(0)} = G
\vspace{-7pt}
\end{equation}
Let $K_{n,n} = (V \cup W, V \times W)$, where $V = W = \{1, 2,\, \cdots\,, n\}$, and $BG =(V \cup W, E) $ be a  subgraph of $ K_{n,n}$, on $2n$ nodes, where $ E \subset V \times W$.

\vspace{-07pt}
\vspace{-4pt}
\subsection{The Mapping: $\mathbf{S_n}$ Generating Set to Perfect Matching Generators}
\vspace{-10pt}
We choose the generating set $K$ of $S_n$ by choosing the set of right representatives $U_i$, $1 \le i < n$,  as transpositions, for the stabilizer chain of subgroups in \eqref{e:tower}, i.e.,\\
\vspace{-4pt}
\begin{equation}\label{e:cosetReps}
U_i = \lbrace I, (i, i+1), (i, i+2), ~\cdots, ~ (i, n)\rbrace, ~ 1 \le i < n.
\vspace{-4pt}
\end{equation}
Then the generating set $K$ of $S_n$ is
\vspace{-4pt}
\begin{equation}\label{e:cosetRep1}
\vspace{-4pt}
K=\bigcup U_i = \lbrace I, (1, 2), (1, 3), ~\cdots, ~(1, n), (2, 3), (2, 4), ~\cdots,
~ (2, n), ~\cdots, ~ (n-1, n)\rbrace
\vspace{-4pt}
\end{equation}
Let
$\mathbb{M}(BG')$ denote the set of permutations realized as perfect matchings in the bipartite graph $BG'$.
Let $BG_i$ denote the sub
(bipartite) graph of $BG=K_{n,n}$, induced by the subgroup $G^{(i)}$, such that $\mathbb{M}(BG_i) = G^{(i)}$.
 \par
A \emph{partition representative}, $g(i)$, derived from $U_i$ (using Theorem 3.1 in \cite{aslam-2017}), $1 \le i \le n$, for $K_{n,n}$,   is defined as:
\begin{equation}\label{e:mGenset}
\vspace{-03pt}
 g(i) \,\defn\, \big\{(i k, t i) \, \mid \, k,t \in \{i+1,\, \cdots\,, n\} \big\}
 \bigcup \big \{(ii, ii) \big\},
\end{equation}
where $(v_i,w_k,v_t,w_i)$ is a cycle of length 4 in $BG_{i-1}$.\par

\par

The following Lemma (3.4 in \cite{aslam-2017}) states the exact mapping between $U_i$ and $ g(i)$.
\par
\begin{lemma} \label{L:mapGenset}
There exists a 1-1 mapping
\vspace{ -7pt}
$$
 h : G^{(i)} \times U_i \longrightarrow g(i) \times M(BG_i),\vspace{ -0pt}
 $$
\par
\vspace{ -8pt}
 \hskip 0.0in s.t., $\forall (\pi,\psi) \in G^{(i)} \times U_i$,
   $\pi\psi $ is realized by a unique pair $ (x_i, pm_i) \in g(i) \times M(BG_i)$ using a unique cycle $(v_i,w_k,v_t,w_i)$ of length 4 in $BG_{i-1}$, defined by $x_i = (ik,ti)$,  such that the edge pair $x_i$ is covered by $\pi\psi$ and the other two alternate edges in the cycle are covered by $\pi$.\\ When $\psi =I$, the identity in $S_n$, the cycle collapses to one edge $x_i =(ii,ii)$ covered by $\pi$ and $\pi\psi$ both.
\end{lemma}

\subsection{The Generating Graph}
\vspace {-9pt}
\begin{wrapfigure}[19]{r}{0.38\textwidth}
\vspace {-0pt}
\vspace {-19pt}
\includegraphics[scale=0.390]{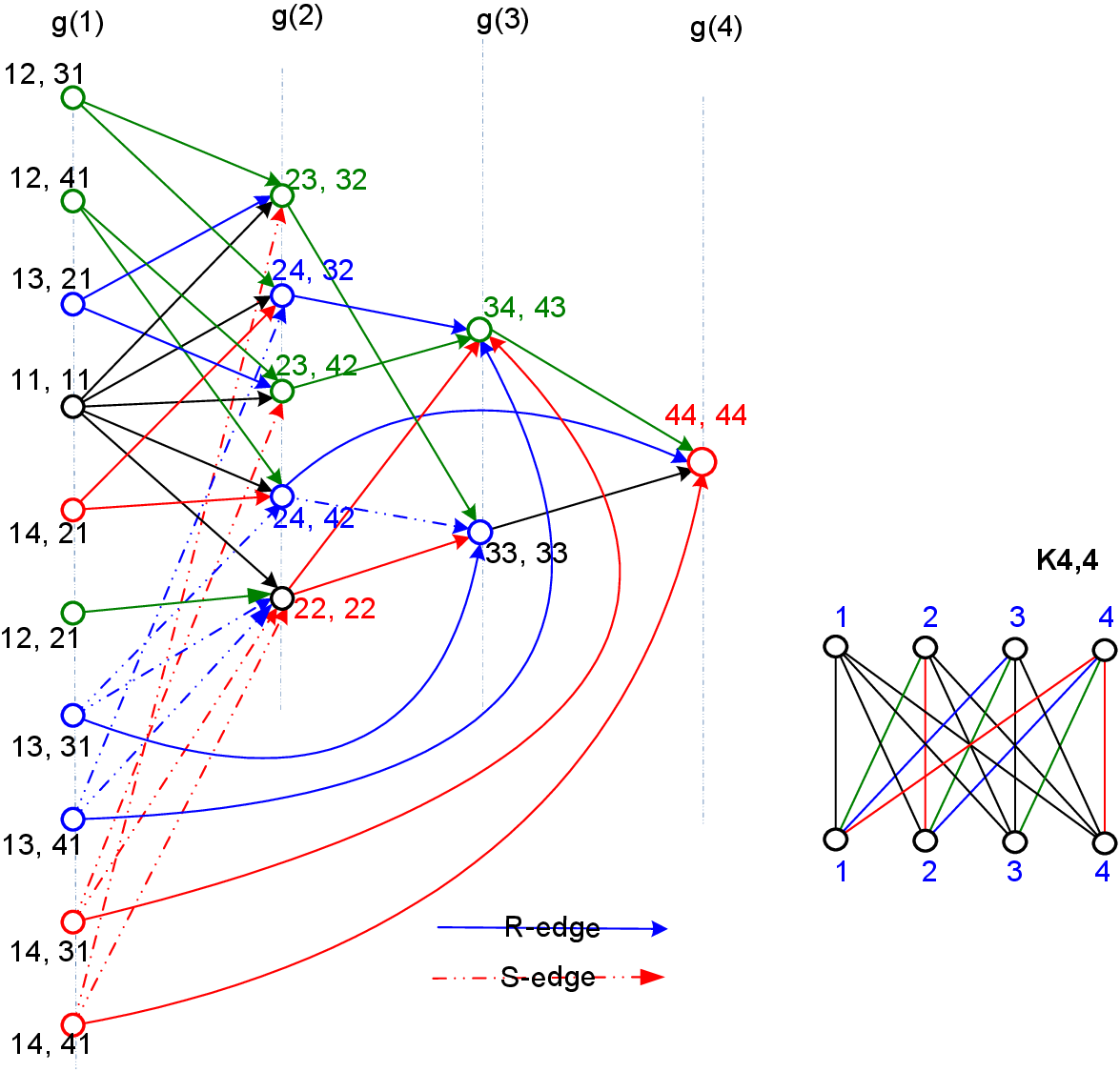}
\caption{\textbf{the Generating Graph} $\Gamma(4)$}\label{FG:GenK44}
\end{wrapfigure}

The generating graph $\Gamma(n)$ for  $K_{n,n}$ models
the two binary relations $R$ and $S$  over $\bigcup g(i)~$ (defined in \cite{aslam-2017}).\\
$\Gamma(n) \defn (V, ~E_R \cup E_S)$, where
$ V = \bigcup g(i)~$,\\
{$E_R = \{ a_i a_j\,|\, a_i R a_j, ~ a_i \in g(i), a_j \in g(j) ~1 \le i < j \le n \}$, and }\\
{$E_S = \{ b_i b_{i+1} \,| \, b_i S b_{i+1}, ~b_i \in g(i) \text{ and }b_{i+1} \in g(i+1),~1 \le i < n \}$.}
\par


\subsubsection*{\textbf{Valid Multiplication Path (\vmp/\cvmp)}}
\vspace{-6pt}
\begin{definition}\label{D:CVMP}
 Let $p= x_i x_{i+1} ~\cdots~ x_{j-1}x_j$ be any path formed by the adjacent $R$- and $S$-edges in $\Gamma(n)$ such that exactly one node $x_r$ is covered in each node partition $r$, where $x_r \in g(r)$, $1 \le i \le r \le j\le n$.
Then $p$ is a \emph{ valid multiplication path} if $\forall (x_r, x_s)$,  on $p$, where $s > r$, we have either $x_r R x_s$ or the edge pairs $x_r $ and $x_s$, are vertex-disjoint in $K_{n,n}$, and  $x_r R x_s$ is false.\par
Further, $p$ is a \emph{Complete Valid Multiplication Path} (CVMP) if for every $R$-edge, $x_r R x_t$, (direct or jump edge) beginning at $x_r$ in $p$, $i \le r < j$, $x_t$ is covered by $p$, i.e., $ r < t \le j$.
\end{definition}
\vspace{-6pt}
\begin{wrapfigure}[0]{r}{\textwidth}
\vspace {-0pt}
\end{wrapfigure}
\vspace {-10pt}
\subsubsection*{General Specification for Multiplying two Nodes}
\vspace{-10pt}
\begin{wrapfigure}[6]{r}{0.38\textwidth}
\vspace{-015pt}
\center {
\includegraphics[scale=0.28]{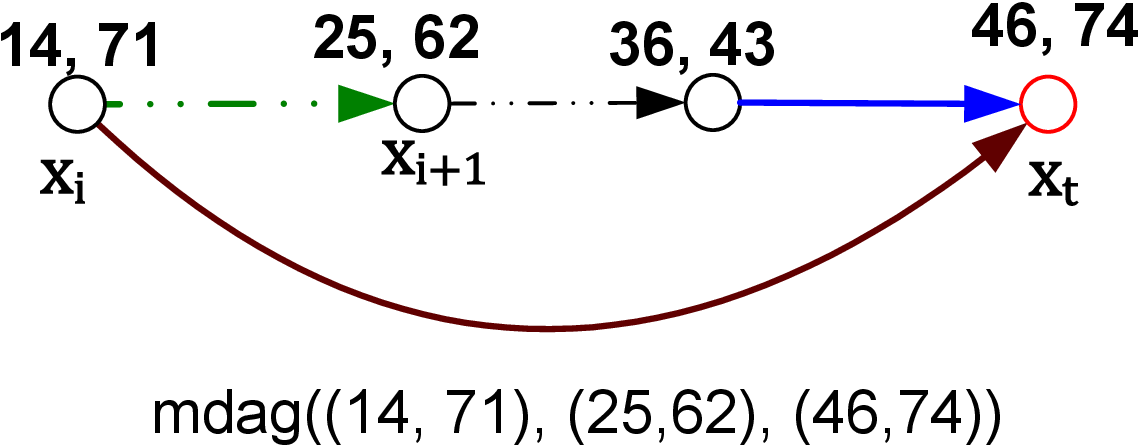}
 \caption{\textbf{A Simple MDAG}}\label{FG:mdagBasic}
 }
\vspace{-0pt}
\end{wrapfigure}

\vspace {-0pt}
A \emph{Multiplying Directed Acyclic Graph} (MDAG), denoted as $mdag(x_i, x_{i+1}, x_t)$, is a general specification for ``multiplying'' two nodes $x_i$ and $x_{i+1}$ in adjacent node partitions in $\Gamma(n)$, where $x_i S x_{i+1}$, and  $x_i R x_t$ defines an  $R$-edge such that all three nodes, $x_i$, $x_{i+1}$ and  $x_t$ are covered by a common \vmp. Clearly, in the  extreme case $mdag(x_i, x_{i+1}, x_t)$ reduces to an $R$-edge defined by $x_i R x_{i+1}$, with $x_{i+1}= x_t$.\\
\begin{wrapfigure}[0]{r}{\textwidth}
\vspace {-0pt}
\end{wrapfigure}

\subsubsection{\textbf{Perfect Matching Represented by a \cvmp}}
\vspace{-8pt}
Let $\psi(x_i)$ denote the transposition $\psi = (i,k) \in S_n$ where $x_i= (ik, ji) \in g(i)$. Let $SE (x_i x_j)$ of an $R$-edge be defined  as in
\eqref{e:SE} later under \ref{ss:ER} Edge Requirements.
\begin{lemma}\label{L:matchingSpecs}
Every $\cvmp$, $p= x_1 x_{2} ~\cdots~ x_{n-1}x_n$ in $\Gamma(n)$, of length $n-1$, where $x_r \in g(r)$, $ r \in [1..n]$ represents a unique
permutation $\pi \in S_n$ realized as a perfect matching $E(\pi)$ in $K_{n,n}$, given by
\begin{subequations}
\vspace{-2pt}
\begin{equation}\label{e:CVMP-perm2}
 E(\pi) = \bigcup_{ x_i \in p} {x_i} - \{SE(x_j x_k)\mid x_jR x_k,\, (x_j, x_k) \in p\}, \text{ and}
\end{equation}
\vspace{-5pt}
\begin{equation} \label{e:CVMP-perm1}
\pi = \psi(x_n) \psi(x_{n-1}) ~\cdots ~ \psi(x_{2}) \psi(x_1),
\end{equation}
\end{subequations}
where $\psi(x_r) \in U_r$ is a transposition defined by the edge pair $x_r$, and $U_r$
 is a set of right coset representatives of the subgroup $G^{(r)}$ in $G^{(r-1)}$ such
 that $U_n \times U _{n-1} \cdots U_2 \times U_1$ generates $S_n$.
\end{lemma}
\vspace {-10pt}
\subsubsection*{Notation: The labeling of nodes and edges in $\Gamma(n)$}
\vspace{-10pt}
\emph{Assuming the nodes in $K_{n,n}$ are labeled from $[0..9]$,
an edge pair $(iv,wi) \in K_{n,n}$ is then labeled as the node $(iv,wi)$ in $\Gamma(n)$, while the $R$-edges $((iv,wi), (wv,tw))$ are labeled by $+wv$.}

 \par
\begin{figure}[h]
\vspace{-10pt}
\hspace{-0.0in}
\center{
  \includegraphics[scale=0.70]{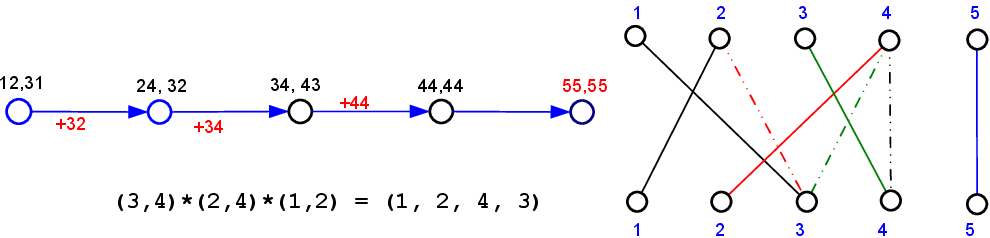}
  \caption{\textbf{A Perfect Matching Composition}}\label{FG:PMcomposition}
  }
\vspace{-7pt}
\end{figure}
\par


\vspace{-21pt}
\subsection{The Partition}
\vspace{-11pt}
The set under the partition is all the $n!$ possible perfect matchings in a bipartite graph. The equivalence classes are the disjoint subsets, induced by the missing edges in each potential perfect matching.

\vspace{-11pt}
\subsubsection{The Edge Requirements}\label{ss:ER}
\vspace{-11pt}
\emph{Edge Requirement} of a CVMP is an algebraic formulation of the perfect matching behavior that every node in the bipartite graph is incident with exactly one edge, i.e., the matched edge.
  \par
  Let $p = x_1 x_2 \cdots x_{n-1} x_n$ be a CVMP in $\Gamma(n)$ for a bipartite graph $BG$.
The Edge Requirement of a node $x_i \in g(i)$ in $p$ is
\begin{equation}
ER(x_i) \defn \{ e~|~ e \in x_i \in g(i) ~\text{ and } e \notin BG \}
\end{equation}
The edge pair $x_i$ represents an initial assignment of the matched edges incident on the node pair $(v_i, w_i)$, in composing a perfect matching.\\
The \emph{surplus edge}, $SE(x_t x_i)$, of an $R$-edge $x_t x_i$
\vspace{-8pt}
\begin{equation}\label{e:SE}
\vspace{ -3pt}
SE(x_t x_i) \defn \text{ the edge $e \in x_t$ covered by the associated $R$-cycle defined by $x_i R x_t$. }
\end{equation}
\vskip -0pt
When the given graph is not a complete bipartite graph, the edge requirement of a node $x_i$ on $p$ can be met by the surplus edge, $SE(x_t x_i)$, as determined by the $R$-edge $x_t x_i$ incident on $x_i$.
For example, in Figure \ref{FG:PMcomposition}, for the CVMP
$p = (12, 31)\cdot(24, 32)\cdot(34, 43)\cdot(44, 44)$, the initial $ER = \{ 44, 34, 32\}$ of various nodes on $p$ is satisfied by the SE of the incident $R$-edges.
%
\par
The Edge Requirement $ER(p)$ of a VMP, $p$ for bipartite graph $BG$, is the collection of each of the nodes' Edge Requirement that is not satisfied by the SE of the $R$-edges incident on that node. That is,
\begin{equation}\label{e:ER-path}
\vspace{ -10pt}
ER(p) \defn \bigcup_{x_i \in p} ER(x_i) - \big( \{ SE(x_j x_k)~| ~x_j, x_k \in p \} \bigcap \big( \bigcup_{x_i \in p} ER(x_i)~\big)~\big)
\vspace{ -0pt}
\end{equation}

In \cite{aslam-2017} Lemma 4.10, we show that a \cvmp~ of length $n-1$ represents a perfect matching in a bipartite graph $BG$ iff $ER(p) =\emptyset$.
\vspace{ -10pt}
\subsubsection{MinSets: The VMPs of Common ER}
\vspace{-8pt}

\begin{figure}[h]
 \center
\includegraphics[scale=0.40]{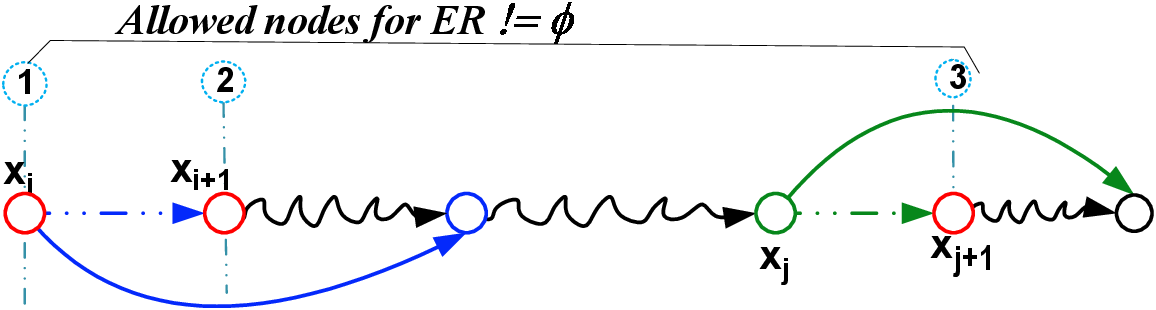}
\vskip 0pt
\caption{\textbf{An Abstract MinSet: $\mathbf {MinSet(mdag\less x_i \more, mdag\less x_j \more)}$}}\label{FG:DefnMinSet} \vspace{-18pt}
\vspace{-5pt}
\end{figure}

Let $mdag\langle x_i \rangle = mdag(x_i, x_{i+1}, x_{r}), r > i+1$, denote a family of mdags.
Let $m_i = mdag\langle x_i \rangle $ at some node $x_i$ in the node partition $i$.\\
Let $ER^p(x_j)$ denote the ER of a node $x_j$ covered by a VMP, $p$.
\vspace{-6pt}
\begin{definition}\label{D:MinSet0}
A $MinSet(m_i, m_{j})$, $1 \le i < j \le n-1$, is the largest subset of
$\vmpset(m_i, \, m_{j})$, where each $p \in MinSet(m_i, \,m_{j})$
has a common ER, $ER(p)$, such that
\vspace{-10pt}
\begin{quote}
\vspace{-15pt}
 \item $\forall (p, x_k) \in MinSet(m_i, \,m_{j})$, the common ER, $ER^p (x_k) =\emptyset$
     except for the 3 common nodes, $x_i, ~~x_{i+1}$, and $x_{j+1}$, in 3 distinguished node partitions ($i, ~{i+1}$, and ${j+1}$) (Fig \ref{FG:DefnMinSet}).
\end{quote}	
\end{definition}
 \subsubsection*{Representation of a MinSet}
\vspace{-10pt}
 A MinSet has a representation similar to that of a VMPSet except for the additional attributes for the common ER and the incident $R$-edges.
\par
$EdgesAtNode ~(\text{ \textbf{Node}, \{incident edges\}})$;\\
$NodePartition \text{ \textbf{Array}[ ] \emph{of }} EdgesAtNode$;\\
\vspace{-8pt}
\begin{equation}
\vspace{-0pt}
\begin{split}
\mathbf{MinSet(mdag\langle x_i \rangle, ~mdag\langle x_j\rangle, ER(x_{j+1}) ) =} \textbf{ Struct } \{\\
&MdagPair \text{ (mdag\less $x_i$\more, ~mdag\less $x_j$\more);}\\
&PartitionList ~\text{\textbf{Array}[$i\,\cdot\cdot\,(j\!+\!1)$]\, \emph{{of}}}\,\text{NodePartition;}\\
&\text{ //ER at 3 distinguished node positions}\\
& CommonER ~ER(x_i);~ ~\\
& CommonER ~ER(x_{i+1});~ \\
& CommonER ~ER(x_{j+1});~ \\
&Count ~\textbf{integer}; \text{// the count of all the contained VMPs} \\
&\}
\end{split}
\vspace{-18pt}
\end{equation}


\vspace{-015pt}
\subsubsection{The Structure of a C\vmpset~Partition}
\vspace{-00pt}
 \subsubsection*{The Covering MinSet- a Subset of CVMPSet}
\vspace{-10pt}

Let $\prod$ denote the product of two or more adjacent MinSets, similar to the product of VMPSets.\\
Let $ I = \{i, j_1, j_2, \cdots,~j_{r-1}\}$ be an index set representing the various node partitions induced by the $ER \ne \emptyset$ nodes in $\vmpset(m_i, m_{t})$ such that $|I| =r$, $1 \le r \le t-i$.
\begin{definition}
A \emph{covering minset}, $CMS_{it} (r)$, represents a subset of $\vmpset(m_i, m_{t})$ by
a sequence of $r$ MinSets for the given $\vmpset(m_i, m_{t})$.
\vspace{6pt}
 That is,
\vskip 0.0in
$CMS_{it} (r) \defn \{MinSet ( m_{i}, m_{j_1}), MinSet ( m_{j_1}, m_{j_2}), ~\cdots, ~MinSet ( m_{j_{r-1}}, m_{t})\},$\\
such that
$$
\hskip -0.8in \prod_{i_j \in I} MinSet ( m_{i_{j}}, m_{i_{j+1}}) ~\subseteq \vmpset(m_i, m_{t}).
$$

\end{definition}
In \cite{aslam-2017} we prove the following algebraic expression of the C\vmpset ~partition shown earlier\\ in Figure \ref{FIG:cosetPartitions}:
\vspace{ -05pt}
\begin{lemma}\label{L:CountofCMS}
 Let $CMS_{in}(r)$ be a MinSet sequence of length $r$ representing a subset of $C\vmpset (m_i, m_{n-1})$, where $1 \le r \le n-2$, $1 \le i \le n-2$.
 Then, for all $i, 1\le i\le n-2$,
 \vspace{-4pt}
 \begin{equation}\label{EQ:CMS-partition}
  C\vmpset (m_i, m_{n-1}) = \biguplus_{\hspace{-1pt}r=1}^{n-2} \hspace{0pt}\prod_{\hspace{03pt}\substack{\cms\hspace{-02pt}_{in}(r),\\ i_j \in I}} \hspace{-15pt}MinSet(m_{i_j}, m_{i_{j+1}})
\vspace{ -15pt}
 \end{equation}
\end{lemma}
\vspace{ -8pt}
\subsubsection*{A Generating Set for the MinSet Sequences}
\vskip -08pt
 Now we define a \emph{generating set}, called $\gms$, for generating the  MinSet Sequences which constitute a partition of the C\vmpset. This is to consolidate the generation of all the MinSets shared by the various CVMPSets through their CMS partitions.

\begin{definition}
 A generating set for the MinSet sequences, $\gms(i,n), ~ 1 \le i \le n-2$, for a bipartite graph on $2n$ nodes is a set of MinSets defined as \\
 \[
 \gms(i,n) \defn \big\{MinSet( m_{r}, m_{s})\, \big |\, (r,s) \in [i\! \,\cdot\cdot ~n\!-\!2] \times [i+1\,\cdot\cdot ~n\!-\!1], ~r <s \big \},
 \]
 where $\{(m_r, m_s)\}$ covers $g(r) \times g(s)$.

  \end{definition}

 The following figure illustrates how exponentially many sequences are partitioned into polynomially many equivalence classes by the prefix MinSets in each $CMS_{in}(r)$. \\
\par
\vskip -00pt
\begin{figure}[h]
 \center
 \includegraphics[scale=0.610]{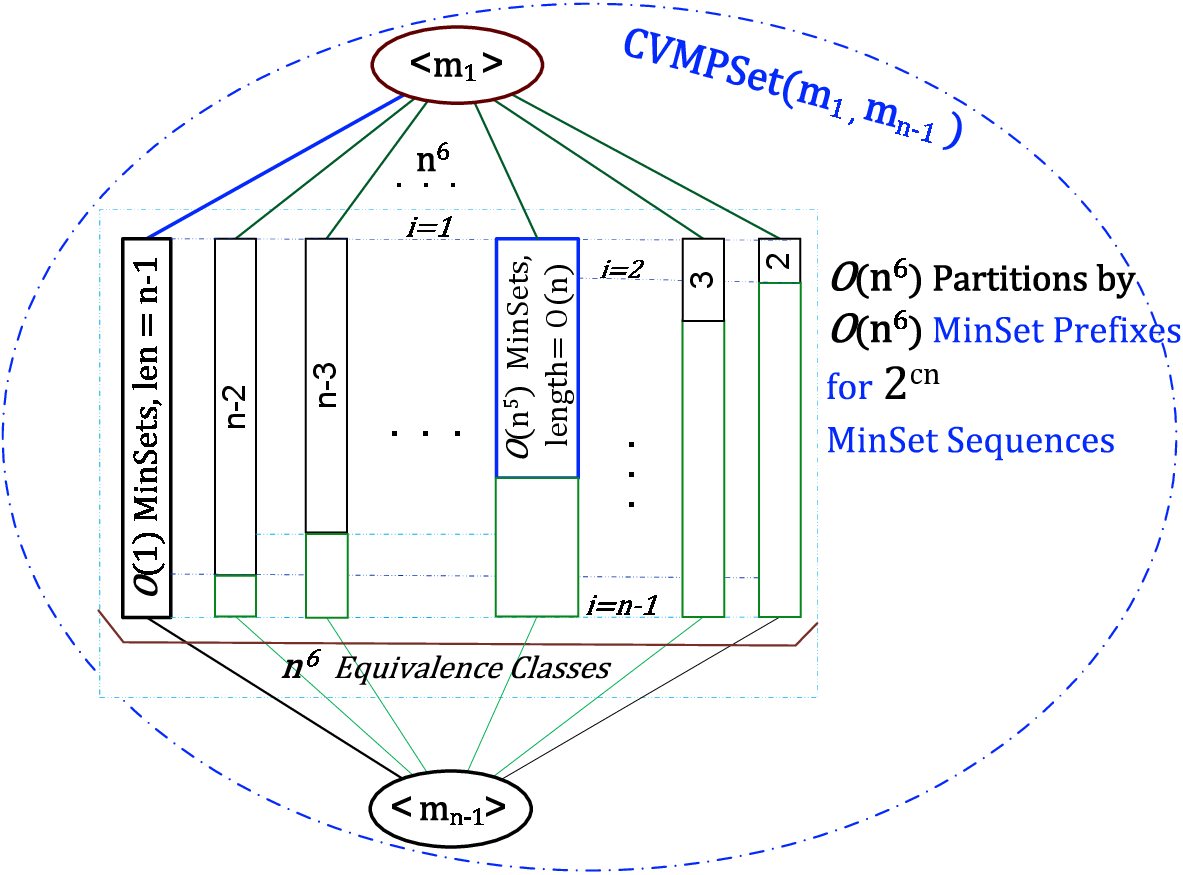}
\vskip 5pt
 \caption{ \textbf{The Final Partition  of a CVMPSet using MinSet Sequences}\label{FIG:PtnHierarchy}}
 \end{figure}
\newpage
\vskip -09pt
\subsection{The Counting Algorithm}\label{S:polybound}
 \vspace{-9pt}


\begin{algorithm}[h]
  \caption{ countPerfectMatchings$( BG)$}
 \begin{algorithmic}\label{ALG:countMatchings}
  \STATE \textbf{Input:} a bipartite graph $BG$ on $2n$ nodes, $n \ge 3$;
  \STATE \textbf{Output:} count of the \emph{perfect matchings} in $BG$;
  \end{algorithmic}
  \hrule
   \begin{description}
  \item[Step 0: \emph{Initialize- Compute the Initial Generating Set of all the MinSet Sequences}]\ \\
   \end{description}
   \vspace{-25pt}
   \begin{algorithmic}[1]
 \STATE $i = n-3$; \COMMENT{ $i$ is the current node partition};
  \STATE \emph{Compute the generating set} $E_M = \{g(r)\,\mid \,1 \le r\le n\}$;
  \STATE \emph{Compute the generating set} $\gms(i+1,n)=\{MinSet(m_{n-2}, m_{n-1})\}$;\COMMENT{the set of all
the MinSet Sequences; each $C\vmpset(m_{n-2}, m_{n-1})$ is a $MinSet\in \gms(n\!-\!2,n)$, with a total count of 6 CVMPs.}
 \end{algorithmic}
 \vspace{5pt}
\hrule
  \begin{description}
  \item[Step 1: \emph{Count}] \ \\
  \hskip +0.35in
  \textbf{if} $(i=0)$ \textbf{then} /\!/ $\gms(1,n)$ may contain the set $\{MinSet(m_1,\, m_{n-1})\} $
 \vspace{-5pt}
 \hspace{0.100in}
 \[ \vspace{-5pt}
\hspace{-0.400in} \text{perfect matching \emph{count} }=\hspace{-10 pt} \sum_{\substack{ER=\emptyset,\\ ( m_1,\, m_{n-1})}} \hspace{-15 pt}MinSet( m_1,\, m_{n-1})\centerdot Count;\ \\
\vspace{-10pt}
\]
\vspace{010pt}
\hskip +0.88in
\textbf{return};\\
\hskip +0.68in
\vspace{-005pt}

\vspace{-10pt}

\item[Step 2: \emph{Increment \& Join the MinSet Sequences}] \ \\
\hskip +01.68in  \hspace{0.350in}$incrementMSS (\gms(i\!+\!1,n))$; /\!/ assuming $n \ge 3$\\
\hskip +0.68in (Follows the structures in Figure \ref{FIG:PtnHierarchy})\\
\text{\emph{decrement} $i$};\\


\vspace{-10pt}
 \vspace{-00pt}\textbf{\hspace{-00pt}repeat} Steps 1-2;\\
 \end{description}\vspace{-15pt}
\hspace{-0.0000in} \textbf{End.}
 \vspace{-00pt}

\vspace{5pt}
  \end{algorithm}
\vspace{5pt}

\subsubsection{The Polynomial Time Bound}
 \vspace{-0pt}
 \begin{claim} \label{CL: PTimebound}
The time complexity of Algorithm \ref{ALG:countMatchings} is $O(n^{45}\log n)$.
\end{claim}
\begin{proof}

 See \cite{aslam-2017}
 \end{proof}

 \subsubsection{Correctness of the Count}
 \vskip -09pt
 \begin{lemma}\label{L: correctCount}
 All the perfect matchings in a bipartite graph $BG$ on $2n$ nodes can be enumerated in polynomial sequential time $O(n^{45}\log n)$.
\end{lemma}
 \vspace{-5pt}
\begin{proof}

The correctness essentially  follows from the  following two assertions as explained in \cite{aslam-2017}:
\vskip -011pt
\begin{enumerate}
\vskip -09pt
	 \item The perfect matching \emph{count} is:
\vskip -010pt
$$
\sum_{\substack{ER=\emptyset,\\ ( m_1, m_{n-1})}} \hspace{-10 pt}MinSet( m_1, m_{n-1})\centerdot Count, ~{\texttt{and}} \vspace{ -010pt}$$
\vskip -12pt
\item All $MinSet( m_1, m_{n-1})$ with $ER=\emptyset$ are contained in $GMS(1,n)$.
\vskip -010pt
\end{enumerate}
Lemma \ref{L:CountofCMS} proves the correctness of the count. Claim \ref{CL: PTimebound} proves the polynomial bound\\ for  Algorithm \ref{ALG:countMatchings}.
\vspace{-5pt}
\end{proof}
 %

\vspace{-10pt}
\section{The Counter-example Re-visited}
\vspace{-6pt}
\subsubsection*{Notation: The labeling of nodes and edges in $\Gamma(n)$}
\vspace{-6pt}
\emph{Assuming the nodes in $K_{n,n}$ are labeled from $N$ using decimal numbers,
a node $(iv,wi) \in \Gamma(n)$ is labeled as $i.v,w.i$, while the $R$-edges $((iv,wi), (wv,tw))$ are labled by $+w.v$, where $``\cdot"$ us used as a delimiter to separate the node labels. When the node numbers are $0,1,2,\, \cdots \,9$, we will ignore this delimiter $``\cdot"$.}

\par
The following figure [Fig. \ref{FG:GroupMultRefuteFig}(b)] shows various MinSets needed to correctly compute the perfect matchings.\\
 The final step (Step 1 of Algorithm 2.1) creates $GMS(1,9)$ with two  MinSet sequences derived from $C\vmpset(m_1, m_8)$ and $C\vmpset(m_1',m_8)$ respectively in the given bipartite graph:
 \begin{enumerate}
\item $\{MinSet(m_1, m_5),~ MinSet( m_5, m_{8})\}$, where $m_1 = mdag(c_1, a_2, c_3), ~m_5 = mdag(a_5, c_6, c_7)$, and
\item $\{ MinSet( {m'_1}, m_{8})\}$, where $m'_1 = mdag(c_1, b_2, c_3)$.
\end{enumerate}
Now the only MinSet sequence, $MinSet( {m'_1}, m_{8})$, in (2) contains C\vmp s of length 8, each having $ER=\emptyset$, giving the perfect matching count as 2.
\par
\vspace{15pt}
\renewcommand{\abovecaptionskip}{+10pt}
\renewcommand{\belowcaptionskip}{-0pt}
\begin{figure}[h]

\center
\includegraphics[scale=0.859]{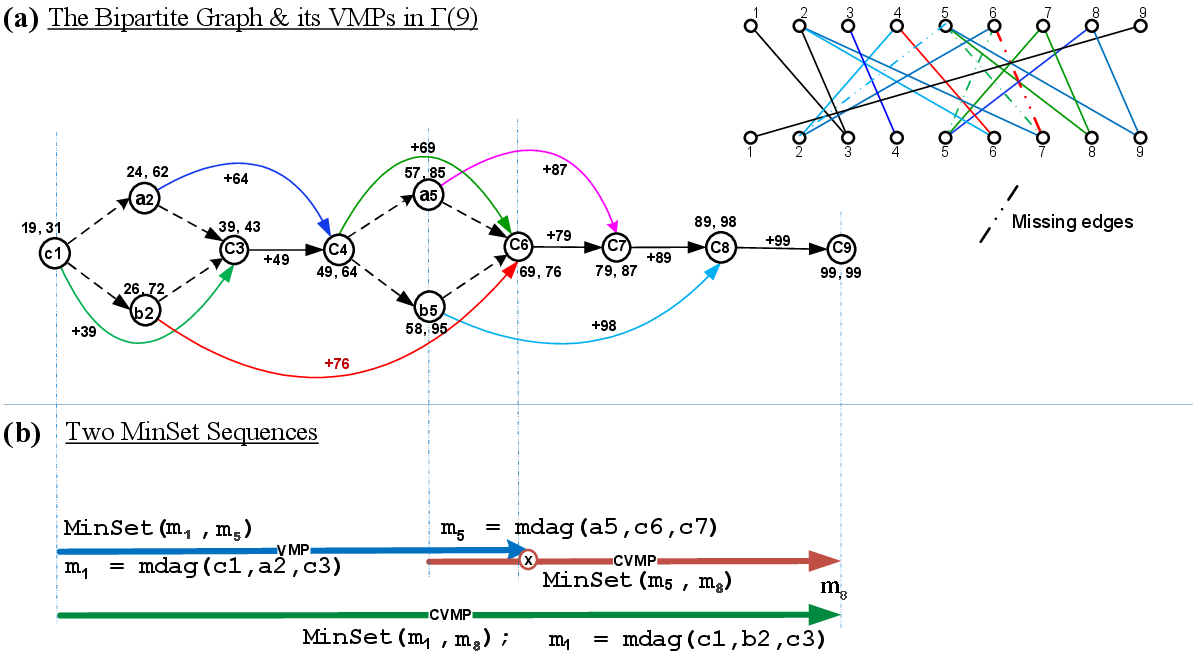}
\vspace{5pt}
\caption{\textbf{Corrected Evaluation of VMPSets}}\label{FG:GroupMultRefuteFig}
\end{figure}

\newpage
\section{Acknowledgement}
\vspace{-5pt}
The author would like to express his sincere gratitude to the authors \cite{Feraro-2009} for finding a logical error in the earlier counting algorithm in \cite{aslam-2008}.

\bibliographystyle{amsalpha}
\bibliography{permalgebraArxiv94}

\end{document}